\documentclass[12pt]{iopart}

\usepackage{iopams} 
\usepackage{setstack}
\usepackage{mathrsfs}
\usepackage{epsfig}
\usepackage{graphicx}
\usepackage{bm}
\usepackage{cite}

\begin{document}

\newcommand{\ped}[1]{_{\mathrm{#1}}}
\newcommand{\api}[1]{^{\mathrm{#1}}}
\newcommand{\diff}[1]{\mathrm{d}#1}
\newcommand{\pped}[1]{_{\scriptscriptstyle #1}}
\newcommand{\aapi}[1]{^{\scriptscriptstyle #1}}
\newcommand{\dfrac}[2]{{\displaystyle\frac{#1}{#2}}}

\title{Prospects for all-optical ultrafast muon acceleration}

\author{F Peano$^1$, J Vieira$^1$, R Mulas$^2$, G Coppa$^2$, R Bingham$^3$ and L O Silva$^1$}
\address{$^1$ GoLP/Institudo de Plasmas e Fus\~ao Nuclear, Instituto Superior T\'ecnico, 1049-001 Lisboa, Portugal}
\address{$^2$ Dipartimento di Energetica, Politecnico di Torino, 10129 Torino, Italy}
\address{$^3$ Space Science \& Technology Dept, Science \& Technology Facilities Council,
Rutherford Appleton Laboratory, Harwell Science and Innovation Campus, Didcot,
Oxon, OX11 0QX, UK}
\eads{\mailto{fabio.peano@ist.utl.pt}, \mailto{luis.silva@ist.utl.pt}}

\begin{abstract}
A scheme for fast, compact, and controllable acceleration of heavy particles in vacuum has been recently proposed [Peano F \textit{et al.} 2008 \textit{New J. Phys.} \textbf{10} 033028], wherein two counterpropagating laser beams with variable frequencies drive a beat-wave structure with variable phase velocity, leading to particle trapping and acceleration. The technique allows for fine control over the energy distribution and the total charge of the accelerated beam, to be obtained via tuning of the frequency variation. Here, the theoretical bases of the acceleration scheme are described, and the possibility of applications to ultrafast muon acceleration and to the prompt extraction of cold-muon beams is discussed.
\end{abstract}


\submitto{\PPCF}
\maketitle

\section{Introduction}
\label{sec:intro}

The recent developments in the ultraintense-laser technology in the infrared (IR) optical regime made compact acceleration of charged particles a central research topic. The main research directions in the field are electron acceleration in laser-driven plasma waves \cite{wakefield_1,wakefield_2,wakefield_3,wakefield_4,wakefield_5} and proton or light-ion acceleration in laser-solid interactions (cf. \cite{ion_accel_0,ion_accel_1,ion_accel_2,Silva,Esirkepov} and, for a review, \cite{Borghesi,Mendonca,Bulanov}).
Typically, these acceleration schemes are indirect, in the sense that the acceleration is provided by laser-driven field structures in plasmas. As an alternative, direct, all-optical acceleration by the electromagnetic (EM) field of ultraintense laser pulses have been considered for electrons \cite{Esarey_1,vacuum_0,vacuum_1,vacuum_2,Hafizi,vacuum_3,vacuum_4,vacuum_5,vacuum_5b,vacuum_6,vacuum_7,vacuum_7b,vacuum_8} and, very recently, also for ions \cite{Salamin_OL,Salamin_PRL,Peano_NJP,Peano_IEEE}.
In \cite{Peano_NJP}, it has been shown that the beating of two counterpropagating EM waves can trap and accelerate heavy particles, such as protons, heavy ions, and muons, even with nonrelativistic radiation intensities (e.g., $\ll 10^{24}$ W/cm$^2$ for proton acceleration with IR lasers), provided that at least one of the EM waves has variable frequency. The method offers the possibility of efficient control over the relevant beam features (mean energy, energy spread, and total accelerated charge), and can be adapted to different sources of charged particles (including beams, and tenuous gases and plasmas), serving either as a pre-accelerator, creating an energetic beam from heavy particles with low energy, or as an energy booster, accelerating particles from a beam with a given energy.
If the necessary technological requirements are met \cite{Peano_NJP,Peano_IEEE}, the technique can represent an alternative for compact production of high-quality ion beams, as well as a unique option for ultrafast acceleration of muons (which could be important to minimize decay losses) during, or immediately after, the muon cooling process \cite{ioniz_cool_1,ioniz_cool_2,ioniz_cool_3,ioniz_cool_4,ioniz_cool_5,ioniz_cool_6,Frictional_1,Frictional_2,Frictional_3} in a muon collider or a neutrino-factory machine \cite{NFMC1,NFMC2,NFMC3,NFMC4,NFMC5,NFMC6,NFMC7,NFMC8}). 

In this Paper, the theoretical bases of the all-optical acceleration scheme are described (Section \ref{sec:theory}), providing estimates for the beam properties and discussing the laser requirements (Section \ref{sec:scalings}), and the possibilities of applications to ultrafast muon acceleration and to the prompt extraction of cold muon beams are investigated (Section \ref{sec:muons}).

\section{Physical mechanism}
\label{sec:theory}

\begin{figure}[!htb]
\centering \epsfig{file=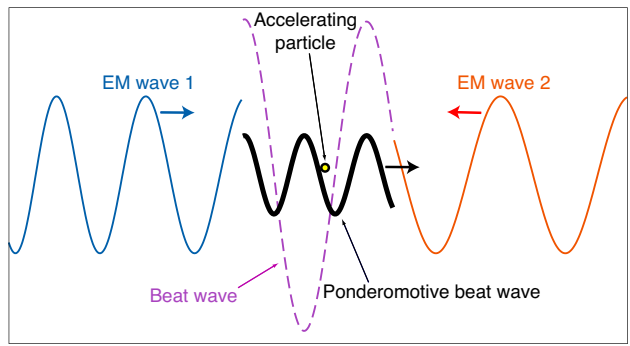, width=3.25in}
\caption{(Color online) Sketch of the acceleration technique: a test particle is trapped and accelerated by the slow ponderomotive beat wave generated by two counterpropagating, variable-frequency lasers. The thin, solid lines represent the vector potential of the two lasers; the thin, dashed line represents the sum of the vector potentials in the superposition region; the thick, solid line represents the ponderomotive beat-wave.}
\label{fig:scheme}
\end{figure}

The acceleration technique presented in \cite{Peano_NJP} exploits two counterpropagating EM waves with variable frequency in order to trap particles in a beat-wave structure with controllable phase velocity, according to the configuration sketched in Fig. \ref{fig:scheme}.
The physical mechanism of acceleration can be analyzed in the framework of a relativistic, one-dimensional, single-particle theory, in which the amplitude of both lasers is constant (two-dimensional particle-in-cell simulations, accounting for spatial charge distributions and finite-size laser pulses, confirmed the validity of the present theory \cite{Peano_NJP,Peano_IEEE}).

\subsection{Field configuration and basic equations}
\label{sec:equations}
In the present theory, the lasers are described as two coherent EM waves (labeled with $j=1,2$) propagating along the $x$ direction, with vector potentials
\begin{equation}
{\bf A}_j\left(x,t\right) = A_j\Big\{\cos\left(\theta\right)\sin\left[\Phi_j\left(\xi_j\right)\right]\hat{{\bf e}}_y + \sin\left(\theta\right)\cos\left[\Phi_j\left(\xi_j\right)\right]\hat{{\bf e}}_z\Big\} \mathrm{,}
\label{eq:vect_pot}
\end{equation}   
where $\hat{{\bf e}}_y$ and $\hat{{\bf e}}_z$ are unit vectors, $\theta$ determines the polarization type (e.g., $\theta = \pi/2$ for linear polarization in the $z$ direction, $\theta = \pi/4$ for circular polarization), $\xi_1 = x - ct$ (laser 1 propagates from left to right), $\xi_2 = -x-ct$ (laser 2 propagates from right to left), $A_j$ is the amplitude, and $\Phi_j\left(\xi_j\right)$ are two arbitrary functions with first derivative $\Phi_j^\prime\left(\xi_j\right)>0$ (henceforth, for the purpose of readability, the arguments $\xi_j$ will be omitted unless necessary). The local wavenumbers, $k_j\left(x,t\right)=\frac{\partial\Phi_j}{\partial x}$, and frequencies, $\omega_j\left(x,t\right)=-\frac{\partial\Phi_j}{\partial t}$, are $k_1\left(x,t\right)=\Phi_1^\prime$, $\omega_1\left(x,t\right)=c\Phi_1^\prime$, $k_2\left(x,t\right)=-\Phi_2^\prime$, and $\omega_2\left(x,t\right)=c\Phi_2^\prime$.
The superposition of the two EM waves can be described in terms of the two corresponding beat waves: (i) a fast, superluminal beat wave, having phase $\Phi\ped{F}(x,t)=\Phi_1+\Phi_2$, wavenumber $K(x,t) = \frac{\partial\Phi\ped{F}}{\partial x} = \frac{1}{2}(\Phi_1^\prime-\Phi_2^\prime)$, frequency $\Omega(x,t) = -\frac{\partial\Phi\ped{F}}{\partial t} = \frac{c}{2}(\Phi_1^\prime+\Phi_2^\prime)$, and phase velocity $V_\phi(x,t)=\frac{\Omega}{K}=c\frac{\Phi_1^\prime+\Phi_2^\prime}{\Phi_1^\prime-\Phi_2^\prime}$, and (ii) a slow, subluminal beat wave, having phase $\Phi\ped{S}(x,t)=\Phi_1-\Phi_2$ wavenumber $k(x,t) = \frac{\partial\Phi\ped{S}}{\partial x} = \frac{1}{2}(\Phi_1^\prime+\Phi_2^\prime)$, frequency $\omega(x,t)= -\frac{\partial\Phi\ped{S}}{\partial t} =\frac{c}{2}(\Phi_1^\prime-\Phi_2^\prime)$ and phase velocity $v_\phi(x,t)=\frac{\omega}{k}=c\frac{\Phi_1^\prime-\Phi_2^\prime}{\Phi_1^\prime+\Phi_2^\prime}$.
The ratio between the frequencies of the two laser beams in $x=0$ at $t=0$ is assumed to be such that the phase velocity of the slow beat wave matches the initial velocity of the particle, $c\beta_0$, i.e., such that $v_\phi(0,0)=c\beta_0$, leading to $\frac{\omega_2(0,0)}{\omega_1(0,0)}=\frac{1-\beta_0}{1+\beta_0}$.

By resorting to the conservation of the transverse canonical momentum, $p_y = -\frac{q}{c}\left({\bf A}_1+{\bf A}_2\right)\cdot\hat{{\bf e}}_y$ and $p_z = -\frac{q}{c}\left({\bf A}_1+{\bf A}_2\right)\cdot\hat{{\bf e}}_z$, the $x$ component of the equation of motion for the particle is written, in cgs units, as 
\begin{equation}
\frac{\diff{p_x}}{\diff{t}} = -\frac{q^2}{2\gamma Mc^2}\frac{\partial}{\partial x}\left({\bf A}_1+{\bf A}_2\right)^2 \mathrm{,}
\label{eq:x_motion}
\end{equation} 
where $q$ and $M$ are the particle charge and mass, respectively, and $\gamma$ is the Lorentz factor, defined as $M^2c^4\gamma^2=M^2c^4+p_x^2c^2+q^2\left({\bf A}_1+{\bf A}_2\right)^2$. 
The squared vector potential in Equation \eref{eq:x_motion} is $({\bf A}_1+{\bf A}_2)^2 = (A_1^2+A_2^2)/2 +  A_1A_2 \cos\left(\Phi\ped{S}\right) + [\sin^2(\theta)-\cos^2(\theta)] \big[ 2A_1A_2 \cos\left(\Phi\ped{F}\right) + \left(A_1^2+A_2^2\right)\cos\left(\Phi\ped{S}\right)\cos\left(\Phi\ped{F}\right)-\left(A_1^2-A_2^2\right)\sin\left(\Phi\ped{S}\right)\sin\left(\Phi\ped{F}\right)\big]/2$.
In the special case of circular polarization, $\sin^2\left(\theta\right)=\cos^2\left(\theta\right)$, $({\bf A}_1+{\bf A}_2)^2=\frac{1}{2}\left(A_1^2+A_2^2\right)+A_1A_2\cos(\Phi\ped{S})$ with arbitrary $A_j$ for both trapped and untrapped particles.
For a generic polarization, the fast terms containing $\Phi\ped{F}$ can be averaged out provided that the two following conditions are met: (i) the variation of $x$ and $p_x$ associated with the transverse quiver motion is negligible, which is verified if $\hat{A}_j = \frac{qA_j}{Mc^2}\ll1$ (here, $\hat{A}_j=\frac{qm}{eM}a_j$, where $m$ and $e$ are the electron mass and the elementary charge, and $a_j=\frac{eA_j}{mc^2}$ is the usual normalized vector potential, corresponding to the peak transverse momentum of an electron in the laser field); (ii) along the particle trajectory $x\ped{p}(t)$, the frequency of the slow beat wave is much lower than the frequency of the fast beat wave, which is automatically satisfied for trapped particles, because in the comoving frame the slow frequency stays always close to zero \cite{Peano_IEEE}. 
Thus, Equation \eref{eq:x_motion} can be averaged over the fast time scale to yield the ponderomotive equation of motion
\begin{equation}
\dfrac{\diff{\hat{p}_x}}{\diff{\hat{t}}} = -\dfrac{\hat{A}_1\hat{A}_2}{2\gamma}\dfrac{\partial}{\partial \hat{x}}\cos\left(\Phi_1-\Phi_2\right) = \dfrac{\hat{A}_1\hat{A}_2}{\gamma}\hat{k}\left(\hat{x},\hat{t}\right)\sin\left(\Phi_1-\Phi_2\right)
\mathrm{,}
\label{eq:x_motion_av}
\end{equation}
where dimensionless quantities (denoted with hatted symbols henceforth) have been adopted: $\hat{k}=\frac{k}{k_0}$, with $k_0=k\left(0,0\right)$, $\hat{t} = k_0ct$, $\hat{x} = k_0x$, and $\hat{p}_x = \frac{p_x}{Mc}$ [in these units, the initial frequencies of the lasers are $\hat{\omega}_1(0,0)=1+\beta_0$ and $\hat{\omega}_2(0,0)=1-\beta_0$]. In Equation \eref{eq:x_motion_av}, the averaged Lorentz factor is defined as $\gamma^2 = 1+\hat{p}_x^2+\hat{A}_1^2 /2+\hat{A}_2^2 /2+\hat{A}_1\hat{A}_2\cos\left(\Phi_1-\Phi_2\right)$ \cite{Mora_Antonsen}.
Equation \eref{eq:x_motion_av} can be obtained from the Hamiltonian $\mathscr{H}\left(\hat{x},\hat{p}_x,\hat{t}\right)=\big[1+\hat{p}_x^2+\hat{A}_1^2 /2+\hat{A}_2^2 /2+\hat{A}_1\hat{A}_2\cos\left(\Phi_1-\Phi_2\right)\big]^{1/2}$, using $\frac{\diff{\hat{p}_x}}{\diff{\hat{t}}}=-\frac{\partial\mathscr{H}}{\partial \hat{x}}$; accordingly, the variation of the particle energy, $\frac{\diff{\gamma}}{\diff{\hat{t}}}=\frac{\partial\mathscr{H}}{\partial \hat{t}}$, is given by
\begin{equation}
\dfrac{\diff{\gamma}}{\diff{\hat{t}}} = \dfrac{\hat{A}_1\hat{A}_2}{2\gamma}\dfrac{\partial}{\partial \hat{t}}\cos\left(\Phi_1-\Phi_2\right) = \dfrac{\hat{A}_1\hat{A}_2}{\gamma}\hat{\omega}\left(\hat{x},\hat{t}\right)\sin\left(\Phi_1-\Phi_2\right)
\mathrm{.}
\label{eq:energy}
\end{equation}
Equation \eref{eq:energy} shows that the presence of a frequency variation is a necessary condition to effectively transfer energy from the beating EM waves to the particle over long times. If the frequency of both lasers is constant (i.e., if $\Phi_1^{\prime}$ and $\Phi_2^{\prime}$ are constants), Equation \eref{eq:energy} reduces to
\begin{equation}
\dfrac{\diff{\gamma}}{\diff{\hat{t}}}=\dfrac{\hat{A}_1\hat{A}_2}{\gamma}\beta_0\sin\left[2\left(\hat{x} - \beta_0\hat{t}\right) + \phi_0\right]
\mathrm{,}
\label{eq:energy_2}
\end{equation}
where $\phi_0 = \Phi_1\left(0\right)-\Phi_2\left(0\right)$ is the initial phase of the beat wave. 
In this situation, $\gamma$ cannot grow over long times (in the frame of reference where $\beta_0=0$, $\mathscr{H}$ is time-indepedent and $\frac{\diff{\gamma}}{\diff{t}}=0$), being bound to oscillate between $(\Gamma\pm\beta_0\hat{P})/(1-\beta_0^2)$, where $\Gamma=[1+\hat{A}_1^2 /2+\hat{A}_2^2 /2+\hat{A}_1\hat{A}_2\cos(\phi_0)]^{1/2}$ $\hat{P}=\{\hat{A}_1\hat{A}_2[1+\cos(\phi_0)]\}^{1/2}$ \cite{Peano_IEEE}. 
However, if at least one of the lasers has variable frequency, causing the phase velocity of the ponderomotive beat wave to vary, and continuous energy transfer from the beating EM waves to the particle is possible.

\subsection{Resonant phase-locked solutions}
\label{sec:resonant}
A general resonant, phase-locked solution of Equation \eref{eq:x_motion_av}, $\hat{X}\left(\hat{t}\right)$, is defined by
\begin{equation}
\Phi_1\left(\hat{X}\left(\hat{t}\right)-\hat{t}\right)-\Phi_2\left(-\hat{X}\left(\hat{t}\right)-\hat{t}\right)=\phi_0
\mathrm{.}
\label{eq:res_cond}
\end{equation}
A particular resonant solution can be found by choosing the expression for $\Phi_1$ (or $\Phi_2$) arbitrarily and then seeking the expression of $\Phi_2$ (or $\Phi_1$) for which Equation \eref{eq:x_motion_av} and condition \eref{eq:res_cond} are simultaneously satisfied. 
Assuming, without loss of generality, that $\Phi_2$ is known, imposing condition \eref{eq:res_cond}, and using $\gamma_{0{\scriptscriptstyle\perp}}^2=1+\hat{A}_1^2/2+\hat{A}_2^2/2+\hat{A}_1\hat{A}_2\cos\left(\phi_0\right)$ and 
$\tau\pped{\parallel}=\gamma_{0{\scriptscriptstyle\perp}}\tau$ (with $\tau$ being the proper time, such that $\frac{\diff{\hat{t}}}{\diff{\tau}}=\gamma$), 
Equations \eref{eq:x_motion_av} and \eref{eq:energy} can be combined  (cf. \cite{Peano_IEEE}) to obtain the evolution equation for $\hat{\xi}_1=\hat{X}-\hat{t}$ as  
\begin{equation}
\dfrac{\diff{^2\hat{\xi}_1}}{\diff{\tau\pped{\parallel}}^2} =\mu_0\Phi_2^\prime\left[-\hat{\xi}_1-2\hat{t}\left(\tau\pped{\parallel}\right)\right]
\mathrm{,}
\label{eq:x_motion_xi}
\end{equation}
where $\mu_0=\hat{A}_1\hat{A}_2\sin\left(\phi_0\right)/\gamma_{0{\scriptscriptstyle\perp}}^2$. Through integration of Equation \eref{eq:x_motion_xi}, one can determine the resonant particle trajectory in the parametric form $\big\{\hat{X}\left(\tau\pped{\parallel}\right),\hat{t}\left(\tau\pped{\parallel}\right)\big\}$ and then use Equation \eref{eq:res_cond} to determine $\Phi_1$. For a generic shape of $\Phi_2$, this process usually requires a numerical approach, because of the intricate dependence of $\hat{t}$ on $\tau\pped{\parallel}$. 
In the special case in which the frequency of laser 2 is constant, i.e., $\Phi_2^\prime=1-\beta_0$, the right-hand side of Equation \eref{eq:x_motion_xi} does not depend on $\hat{t}$ and the calculation can be performed analytically \cite{Peano_IEEE}: integration of Equation \eref{eq:x_motion_xi} yields $\hat\xi_1\left(\tau\pped{\parallel}\right) = \frac{c_0}{2}\left(\mu \tau\pped{\parallel}-2\right)\tau\pped{\parallel}$, where $\mu = \mu_0/\gamma_{0{\scriptscriptstyle\parallel}}$, with $\gamma_{0{\scriptscriptstyle\parallel}}=\sqrt{1-\beta_0^2}$, and $c_0 = \gamma_{0{\scriptscriptstyle\parallel}}\left(1-\beta_0\right) = \sqrt{1-\beta_0}/\sqrt{1+\beta_0}$, leading to 
$\hat{X}\left(\tau\pped{\parallel}\right)=\frac{1}{4\mu c_0}\big[c_0^2\left(\mu\tau\pped{\parallel}-2\right)\mu\tau\pped{\parallel}-2\log\left(1-\mu\tau\pped{\parallel}\right)\big]$,
$\hat{t}\left(\tau\pped{\parallel}\right)=-\frac{1}{4\mu c_0}\big[c_0^2\left(\mu\tau\pped{\parallel}-2\right)\mu\tau\pped{\parallel}+2\log\left(1-\mu\tau\pped{\parallel}\right)\big]$,
$\hat{p}\pped{\parallel}\left(\tau\pped{\parallel}\right)=\frac{\diff{\hat{X}}}{\diff{\tau\pped{\parallel}}}=\big[1-c_0^2\left(1-\mu\tau\pped{\parallel}\right)^2\big]/\big[2c_0\left(1-\mu\tau\pped{\parallel}\right)\big]$, and 
$\gamma\pped{\parallel}\left(\tau\pped{\parallel}\right)=\frac{\diff{\hat{t}}}{\diff{\tau\pped{\parallel}}}=\big[1+c_0^2\left(1-\mu\tau\pped{\parallel}\right)^2\big]/\big[2c_0\left(1-\mu\tau\pped{\parallel}\right)\big]$. 
Finally, by noticing that $1-\mu\tau\pped{\parallel}=\big(1+\frac{2\mu}{c_0}\hat{\xi}_1\big)^{1/2}$, $\hat{X}$ and $\hat{t}$ can be written as functions of $\hat{\xi}_1$ and replaced in Equation \eref{eq:res_cond}, yielding
\begin{equation}
\Phi_1\left(\hat{\xi}_1\right)=\phi_0+\frac{1}{2\mu_0}\log\left(1+2\mu_0\left(1+\beta_0\right)\hat{\xi}_1\right)
\mathrm{.}
\label{eq:res_cond_xi}
\end{equation}

It is useful to analyze the behavior of the resonant solution in the early stage of the acceleration, $\hat{t}\ll\mu^{-1}$, and in the asymptotic limit, $\hat{t}\gg\mu^{-1}$. For $\hat{t}\ll\mu^{-1}$, the above expression for $\hat{t}\big(\tau\pped{\parallel}\big)$ gives, to second order in $\hat{t}$, $\tau\pped{\parallel}\big(\hat{t}\big)\approx \frac{1}{\gamma_{0{\scriptscriptstyle\parallel}}}\hat{t}-\frac{\beta_0}{2\gamma_{0{\scriptscriptstyle\parallel}}^2}\mu\hat{t}^2$ and the explicit dependence of $\hat{p}\pped{\parallel}$, $\gamma\pped{\parallel}$, and $\hat{X}\pped{\parallel}$ on $\hat{t}$ is written as $\hat{p}\pped{\parallel}\big(\hat{t}\big) \approx \gamma_{0{\scriptscriptstyle\parallel}}\beta_0+\mu\hat{t}+\frac{1}{2\gamma_{0{\scriptscriptstyle\parallel}}}\mu^2\hat{t}^2$, $\gamma\pped{\parallel}\big(\hat{t}\big) \approx \gamma_{0{\scriptscriptstyle\parallel}}+\beta_0\mu\hat{t}+\frac{1+\gamma_{0{\scriptscriptstyle\parallel}}\beta_0}{2\gamma_{0{\scriptscriptstyle\parallel}}}\mu^2\hat{t}^2$, and $\hat{X}\big(\hat{t}\big) \approx \beta_0\hat{t} + \frac{1}{2\gamma_{0{\scriptscriptstyle\parallel}}^3}\mu\hat{t}^2$. In the nonrelativistic limit ($\beta_0 \rightarrow 0$, $\gamma_{0{\scriptscriptstyle\parallel}} \rightarrow 1$), the particle accelerates uniformly and its energy grows quadratically in time as $\frac{1}{2}\mu_0^2\hat{t}^2$. For $\hat{t}\gg\mu^{-1}$, the behavior of the solution depends on whether laser 1 and the particle are copropagating ($\mu>0$) or counterpropagating ($\mu<0$): if $\mu>0$, $\tau\pped{\parallel}\big(\hat{t}\big) \approx \frac{1}{\mu} \big[1-\exp\big(\frac{c_0^2}{2}-2c_0\mu \hat{t}\big)\big]$ and, consequently, $\hat{X}\big(\hat{t}\big) \approx \hat{t} - \frac{c_0}{2\mu}$ and $\gamma\pped{\parallel}\big(\hat{t}\big) \approx \hat{p}\pped{\parallel}\big(\hat{t}\big) \approx \frac{1}{2c_0}\exp\big(2c_0\mu \hat{t}-\frac{c_0^2}{2}\big)$; if $\mu<0$, $\tau\pped{\parallel}\big(\hat{t}\big) \approx 2\big(\hat{t}/|\mu c_0|\big)^{1/2}$ and, consequently, $\hat{X}\big(\hat{t}\big) \approx -\hat{t}$ and $\gamma\pped{\parallel}\big(\hat{t}\big) \approx \hat{p}\pped{\parallel}\big(\hat{t}\big) \approx \big(|\mu c_0|\hat{t}\big)^{1/2}$. The behavior is asymmetric because, if $\mu>0$, the frequency must be increased to maintain phase-locking, causing a continuous increase of the ponderomotive force, whereas, if $\mu<0$, the frequency must be decreased, causing a continuous decrease of the ponderomotive force.

\subsection{Particle trapping}
\label{sec:trapping}
Effective particle acceleration can occur also when $\Phi_1$ and $\Phi_2$ do not match the resonant solution exactly. In fact, it is sufficient to choose $\Phi_1$ and $\Phi_2$ appropriately in order to guarantee that the particle trajectory stays close to the beat-wave trajectory until reaching the desired energy, i.e., that the particle is trapped by the beat wave.
For given $\Phi_1$ and $\Phi_2$, the beat-wave trajectory $\hat{x}_{\phi_0}\left(\hat{t}\right)$ (namely, the trajectory of the point of the beat wave having constant phase $\phi_0$) is determined by solving the equation $\Phi_1\left(\hat{x}_{\phi_0}-\hat{t}\right)-\Phi_2\left(-\hat{x}_{\phi_0}-\hat{t}\right)=\phi_0$ with respect to $\hat{x}_{\phi_0}$ [for resonant trajectories, $\hat{X}\left(\hat{t}\right)=\hat{x}_{\phi_0}\left(\hat{t}\right)$]. In order to determine the conditions for the occurrence of trapping, Equation \eref{eq:x_motion_av} is expressed in terms of the phase difference between particle and wave, $\psi=2\big[\hat{x}-\hat{x}_{\phi_0}\left(\hat{t}\right)\big]$, and using the proper time $\tau$, as
\begin{equation}
\frac{\diff^2{\psi}}{\diff{\tau^2}} = -\frac{\partial}{\partial\psi}U\left(\psi,\tau\right)
\mathrm{,}
\label{eq:dpsi_dtau}
\end{equation} 
with $U\left(\psi,\tau\right) = 2\hat{A}_1\hat{A}_2 \cos\big\{ \Phi_1\big[\hat{\xi}_{1,\phi_0}\left(\tau\right)+\frac{\psi}{2}\big]-\Phi_2\big[\hat{\xi}_{2,\phi_0}\left(\tau\right)-\frac{\psi}{2}\big]\big\}+ 2\alpha_{\phi_0}\left(\tau\right)\psi$, 
where $\hat{\xi}_{1,\phi_0}$, $\hat{\xi}_{2,\phi_0}$, and $\alpha_{\phi_0}$ depend on $\tau$ through $\hat{t}$ as $\hat{\xi}_{1\phi_0}\left(\tau\right)=\hat{x}_{\phi_0}\left(\hat{t}\right)-\hat{t}$, $\hat{\xi}_{2,\phi_0}\left(\tau\right)=-\hat{x}_{\phi_0}\left(\hat{t}\right)-\hat{t}$, and $\alpha_{\phi_0}\left(\tau\right)=\frac{\diff^2}{\diff{\tau^2}}\hat{x}_{\phi_0}\left(\hat{t}\right)=\gamma\left(\hat{t}\right)\frac{\diff{}}{\diff{\hat{t}}}\left[\gamma\left(\hat{t}\right)\frac{\diff{}}{\diff{\hat{t}}}\hat{x}_{\phi_0}\left(\hat{t}\right)\right]$. According to Equation \eref{eq:dpsi_dtau}, trapping is allowed only if the frequency variation is slow enough to guarantee that the effective potential $U$ presents local minima.
\begin{figure}[!htb]
\centering \epsfig{file=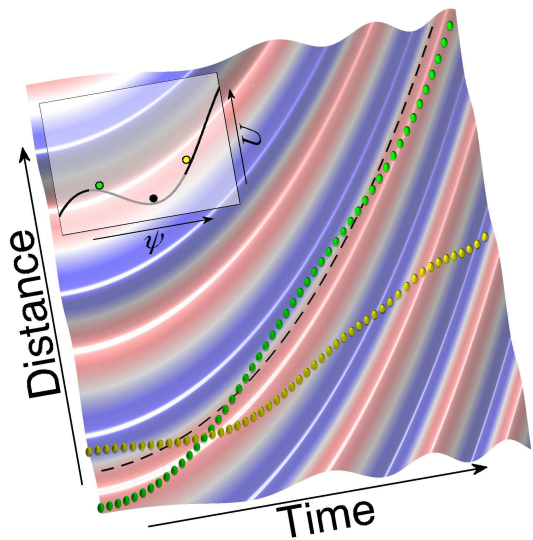, width=3in}
\caption{(Color online) Space-time evolution of the dimensionless ponderomotive force $-\frac{\partial}{\partial{\hat{x}}}\cos\left(\Phi_1-\Phi_2\right)$ (color scale, red and blue denoting accelerating and decelerating regions, respectively), and trajectories of a resonant particle (dashed line), a non-resonant, trapped particle (green spheres), and a non-resonant, untrapped particle (yellow spheres), for a situation in which $\beta_0=0$, $\Phi_2^\prime = 1$, and $\Phi_1$ is given by expression \eref{eq:res_cond_xi}, with $\hat{A}_1\hat{A}_2=5\times10^{-4}$ and $\phi_0=5\pi/6$. The inset represents the effective potential $U$ at $\hat{t}=0$, gray marking the trapping region, and the corresponding initial positions of the three particles.}
\label{fig:trajectories}
\end{figure}
For trapped particles, $|\psi|\ll 2\hat{\xi}_{j\phi_0}$ and $U$ can be approximated as $
U\left(\psi,\tau\right) \approx 2\hat{A}_1\hat{A}_2\cos\big\{\hat{k}\big[\hat{x}_{\phi_0}\left(\hat{t}\right),\hat{t}\big]\psi +\phi_0\big\} + 2\alpha_{\phi_0}\left(\hat{t}\right)\psi$    
leading to the necessary condition for particle trapping
\begin{equation}
\left|\alpha_{\phi_0}\left(\hat{t}\right)\right|<\alpha\pped{\mathrm{M}}\left(\hat{t}\right)=\hat{A}_1\hat{A}_2\hat{k}\left[\hat{x}_{\phi_0}\left(\hat{t}\right),\hat{t}\right]
\mathrm{,}
\label{eq:trap_cond}
\end{equation}
where $\alpha\pped{\mathrm{M}}\left(\hat{t}\right)$ is the maximum value of the ponderomotive force. Thus, trapping is allowed only if the maximum ponderomotive force of the beat wave is greater than the inertial force associated with the beat-wave acceleration.
The particular situation in which $\left|\alpha_{\phi_0}\left(\hat{t}\right)\right|=\alpha\pped{\mathrm{M}}\left(\hat{t}\right)$ corresponds to a resonant solution in the particular case $\phi_0 = \pm \pi/2$. When $\Phi_1$ and $\Phi_2$ correspond to a resonant solution and $|\phi_0| \neq \pi/2$, regions where trapping is possible always exist, although the solution is stable for $\cos\left(\phi_0\right)<0$ and unstable for $\cos\left(\phi_0\right)>0$, with $\psi=0$ corresponding to the bottom and top of a potential well, respectively. 
The relative width of the potential well in $U\left(\psi,0\right)$, $\Delta\psi$, can be used to estimate the trapping efficiency $\eta $ (defined as the fraction of trapped particles) as $\eta \approx\frac{\Delta\psi}{2\pi}\simeq 1-\frac{2}{\pi}\arcsin[\alpha_{\phi_0}\left(0\right)/\alpha\pped{\mathrm{M}}]$.
Examples of typical trapped and untrapped trajectories are shown in Fig. \ref{fig:trajectories}.

\subsection{Acceleration with linearly chirped lasers}
\label{eq:linear_chirp}
\begin{figure}[!htbp]
\centering \epsfig{file=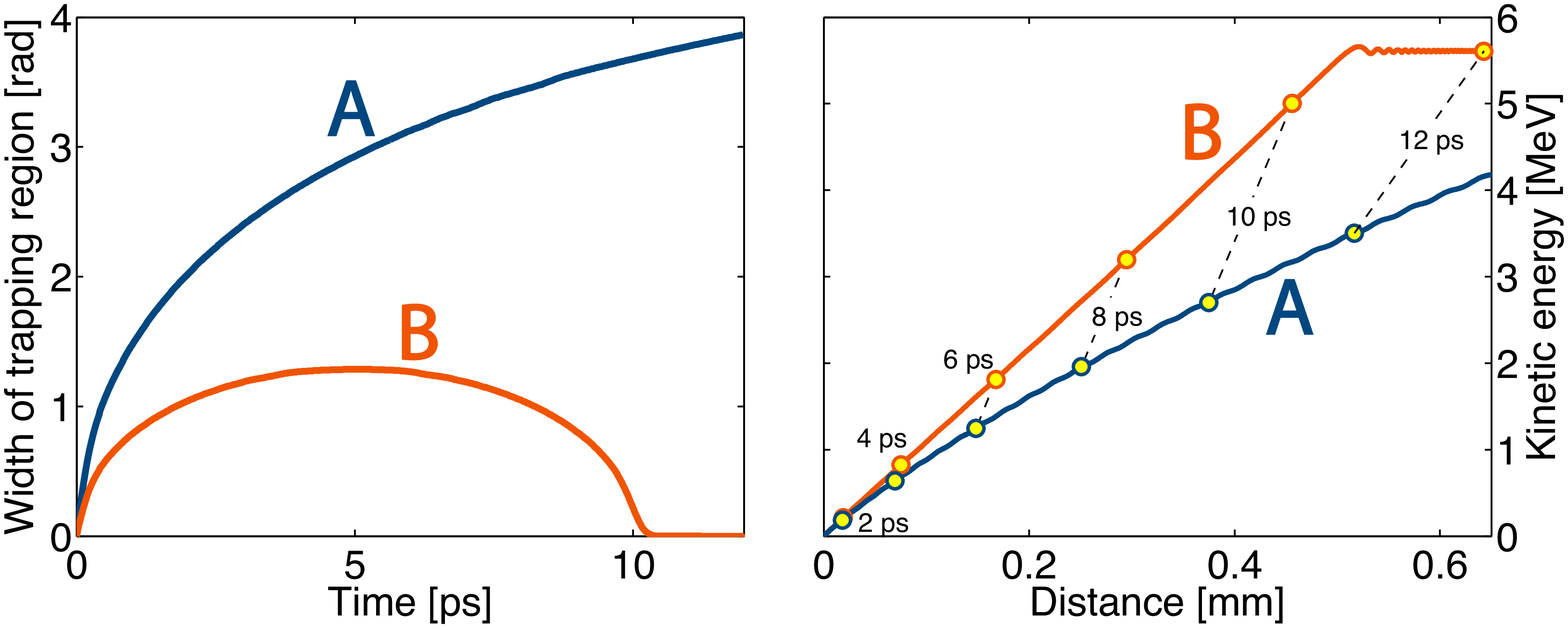, width=4in}
\caption{(Left panel) Width of trapping region as a function of time and (right panel) evolution of particle energy as a function of the acceleration distance when only laser one is chirped with $\sigma_1 = -A_1A_2$ (case A), and when both lasers are chirped with $\sigma_1 = -0.65A_1A_2$ and $\sigma_2=0.35A_1A_2$ (case B). Particle type: muon. Initial particle velocity: $\beta_0 = 0$. Initial laser intensities: $I_1=I_2=1.3\times10^{18}\mathrm{ W}/\mathrm{cm}^2$. Initial laser wavelenghts: 800 nm.}
\label{fig:linear}
\end{figure}

An important case of variable-frequency lasers, particularly relevant in experiments, is that of linearly chirped beams, in which $\Phi_1$ and $\Phi_2$ take the form $\Phi_1\big(\hat{\xi}_1\big)=\phi_{01}+\big(1+\beta_0\big)\hat{\xi}_1+\sigma_1\hat{\xi}_1^2$ and $\Phi_2\big(\hat{\xi}_2\big)=\phi_{02}+\big(1-\beta_0\big)\hat{\xi}_2+\sigma_2\hat{\xi}_2^2$, where $\phi_{0j}$ are constant phases and $\sigma_j$ are the chirp coefficients. The local frequencies and wavenumbers are given by $\hat{\omega}_1\big(\hat{x},\hat{t}\big)=\hat{k}_1\big(\hat{x},\hat{t}\big)=1+\beta_0+2\sigma_1\hat{\xi}_1$ and $\hat{\omega}_2\big(\hat{x},\hat{t}\big)=-\hat{k}_2\big(\hat{x},\hat{t}\big)=1-\beta_0+2\sigma_2\hat{\xi}_2$. The wavenumber and the frequency of the slow beat wave are $\hat{k}\big(\hat{x},\hat{t}\big)=1+\sigma\pped{-}\hat{x}-\sigma\pped{+}\hat{t}$ and $\hat{\omega}\big(\hat{x},\hat{t}\big)=\beta_0+\sigma\pped{+}\hat{x}-\sigma\pped{-}\hat{t}$, where $\sigma\pped{-}=\sigma_1-\sigma_2$ and $\sigma\pped{+}=\sigma_1+\sigma_2$. The phase of the ponderomotive beat wave is $\Phi_1-\Phi_2=\phi_0+2\big(\hat{x}-\beta_0\hat{t}-\sigma\pped{+}\hat{x}\hat{t}\big)+\sigma\pped{-}\big(\hat{x}^2+\hat{t}^2\big)$, with $\phi_0 = \phi_{01}-\phi_{02}$. Consequently, the beat-wave trajectory is $\hat{x}_{\phi_0}\big(\hat{t}\big)=\frac{\sigma\pped{+}}{\sigma\pped{-}}\hat{t}-\frac{1}{\sigma\pped{-}}\big\{1-\big[1-2\big(\sigma\pped{+}-\beta_0\sigma\pped{-}\big)\hat{t}+\big(\sigma\pped{+}^2-\sigma\pped{-}^2\big)\hat{t}^2\big]^{1/2}\big\}$. 
If only one laser is chirped, i.e., if $\sigma_2=0$ or $\sigma_1=0$, one finds
$\hat{x}^{(1)}_{\phi_0}\big(\hat{t}\big)=\hat{t}-\frac{1}{\sigma_1}\big\{1-\big[1-2\sigma_1\big(1-\beta_0\big)\hat{t}\big]^{1/2}\big\}$ and
$\hat{x}^{(2)}_{\phi_0}\big(\hat{t}\big)=-\hat{t}+\frac{1}{\sigma_2}\big\{1-\big[1-2\sigma_2\big(1+\beta_0\big)\hat{t}\big]^{1/2}\big\}$,
respectively. This suggests that the acceleration can be improved by chirping both lasers, with $\sigma_1\sigma_2<0$. 
However, when the particle and the chirped laser are copropagating (e.g., $\sigma_1<0$ and $\sigma_2=0$), $\alpha_{\phi_0}\left(\hat{t}\right)$ is monotonically decreasing in time, causing the trapping regions to widen \cite{Peano_IEEE}; on the contrary, when the particle and the chirped laser are counterpropagating (e.g., $\sigma_1=0$ and $\sigma_2>0$), $\alpha_{\phi_0}\left(\hat{t}\right)$ is monotonically increasing in time, causing the trapping regions to narrow.  
When chirping both lasers ($\sigma_1 \neq 0$ and $\sigma_2 \neq 0$), the use of a stronger chirp in the laser copropagating with the particle is in general convenient: in such a situation, the general expression for $\alpha_{\phi_0}\left(\hat{t}\right)$ has a minimum for a given time (which is $\hat{t}=0$ if $\sigma_1=-\sigma_2$), such that the width of the trapping regions increases until $\alpha_{\phi_0}\left(\hat{t}\right)$  reaches its minimum, and starts decreasing immediately after.
This is illustrated in Fig. \ref{fig:linear}, showing the trapping and acceleration process for a case wherein only one laser is chirped (case A) and for a case wherein both lasers are chirped (case B): in case A, the amplitude of the trapping region, $\Delta\psi$, grows monotonically towards its maximum value, $\Delta\psi=2\pi$, and the acceleration continues indefinitely; in case B, $\Delta\psi$ reaches a maximum and then starts decreasing until synchronization with the beat wave is lost and the acceleration stops.

\section{Beam properties and laser requirements}
\label{sec:scalings}

Basic estimates for the principal acceleration features are readily obtained in the nonrelativistic regime, wherein $\Phi_j$ can always be approximated as $\Phi_j\big(\hat{\xi}_j\big) \approx \phi_{0j}+\hat{\xi}_j+\sigma_j\hat{\xi}_j^2$, with linear-chirp coefficients $\sigma_j=\frac{1}{2}\Phi_j^{\prime \prime}\big(\xi_j\big)$. To leading order in $\hat{t}$, this leads to the beat-wave trajectory $\hat{x}_{\phi_0}\big(\hat{t}\big)\approx\frac{1}{2}\sigma\pped{-}\hat{t}^2$. Hence, the momentum and kinetic energy of trapped particles can be approximately expressed as $\sigma\pped{-}\hat{t}$ and $\frac{1}{2}\sigma\pped{-}^2\hat{t}^2$, respectively.
The maximum energy gain, $\Delta \epsilon_{\scriptscriptstyle\mathrm{M}}$, which occurs when using the maximum chirp that allows for trapping, $|\sigma\pped{-}|=\hat{A}_1\hat{A}_2$, can then be estimated as $\Delta \epsilon_{\scriptscriptstyle\mathrm{M}}[\mathrm{MeV}]\approx 0.8$ $q^4[e]$ $M^{-3}$[amu] $ I_1[10^{20}\mathrm{W/cm^2}]$ $I_2[10^{20}\mathrm{W/cm^2}]$ $\lambda_0^2[\mu\mathrm{m}]$ $\Delta T^2$[ps], where $\lambda_0=2\pi/k(0,0)$ is the initial laser wavelength; the corresponding acceleration distance scales as $\Delta x[\mu\mathrm{m}] \approx 6$ $q^2[e]$ $M^{-2}$[amu] $I_1^{1/2}[10^{20}\mathrm{W/cm^2}]$ $I_2^{1/2}[10^{20}\mathrm{W/cm^2}]$ $\lambda_0[\mu\mathrm{m}]$ $\Delta T^2$[ps].
According to these estimates, if sufficiently intense lasers are employed, the method can provide ultrafast acceleration of heavy particles over short distances.

The maximum energy gain can also be expressed in terms of the laser-pulse energies as $\Delta \mathcal{E}\pped{M}[\mathrm{MeV}]$ $\approx$ $0.08$ $q^4[\mathrm{e}] M^{-3}[\mathrm{amu}]$$\mathcal{E}_1[\mathrm{J}]\mathcal{E}_2[\mathrm{J}]$$\lambda_{0}^{2}[\mu\mathrm{m}]$$W_{0}^{-4}[\mu\mathrm{m}]$. 
Hence, if $W_{0}$ and $\lambda_{0}$ are given, $\Delta \mathcal{E}\pped{M}$ depends only on the energy of the two laser pulses, and the combination of pulse lengths and intensities can be tuned in order to obtain the desired balance between the relative width of the final energy spectrum (which scales as $\frac{A_1A_2}{\Delta \mathcal{E}\pped{M}}$) and the total accelerated charge (which scales as $A_1A_2$ if the laser durations are longer than the diffraction length, being constant otherwise).
Provided that the available laser energy is sufficient, the other fundamental requirement of the acceleration method is the frequency bandwidth. The frequency excursions, $\Delta \omega_j$, necessary to accelerate a particle from the initial velocity $\beta_0$ to the final velocity $\beta$ are determined by 
\begin{equation}
\beta = \frac{2\beta_0+\Delta\omega_1-\Delta\omega_2}{2+\Delta\omega_1+\Delta\omega_2} \mathrm{.}
\label{eq:domega}
\end{equation}
When only one laser is chirped, $\Delta \omega_j$ are given by $\Delta \omega_1=2\left(\beta-\beta_0\right)/\left(1-\beta\right)$ and $\Delta \omega_2=2\left(\beta-\beta_0\right)/\left(1+\beta\right)$. The minimum frequency excursion per laser is obtained when $\Delta \omega_1=-\Delta \omega_2=\beta-\beta_0$.

The present acceleration technique remains effective also in more general situations involving distributed particle sources and finite-size laser pulses, provided that the particle density, $n$, is sufficiently low to prevent significant distortions of the EM beat-wave structure by the space-charge field, and the longitudinal and transverse variations in the radiation intensity are sufficiently smooth to guarantee the validity of the plane wave approximation in the focal region (as confirmed by two-dimensional particle-in-cell simulations \cite{Peano_NJP,Peano_IEEE}, performed with the Osiris 2.0 simulation framework \cite{osiris}). These considerations lead to the constraint on the particle density $n[10^{19}\mathrm{cm}^{-3}] \ll q^{-1}[e]I_1^{1/2}[10^{20}\mathrm{W/cm^2}]I_2^{1/2}[10^{20}\mathrm{W/cm^2}]$, and, to the estimate for the maximum trapped charge $Q\pped{M}\mathrm{[pC]} \approx 16 \eta q[e]$ $n[10^{19}\mathrm{cm}^{-3}]$ $W_0^4[\mu\mathrm{m}]/\lambda_0[\mu\mathrm{m}]$, where $W_0$ is the overlapped spot size of the lasers and $\eta$ is the fraction of trapped particles. 

\section{Ultrafast muon acceleration}
\label{sec:muons}
\begin{figure}[!htbp]
\centering \epsfig{file=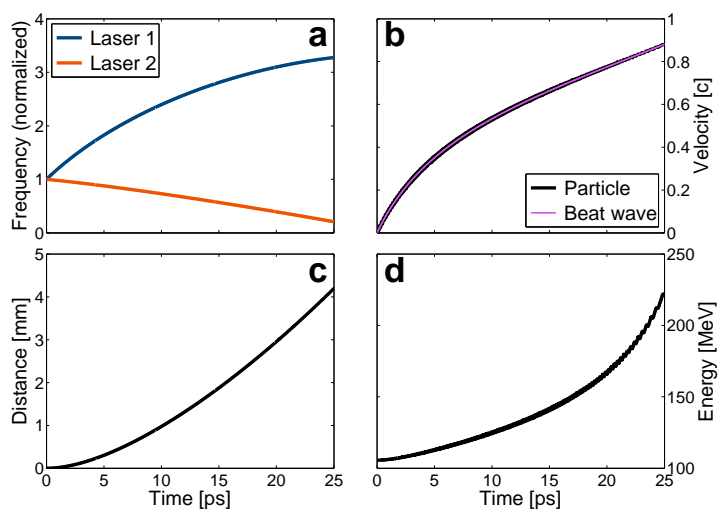, width=3.75in}
\caption{Example of single-stage muon acceleration: (a) laser frequencies along the muon trajectory, (b) muon velocity and beat-wave phase velocity, (c) muon trajectory, and (d) evolution of the muon energy. Laser parameters: $I_1=I_2=1.4\times10^{19} \ \mathrm{W}/\mathrm{cm}^2$, $\sigma_1=-4.4\times10^{-5}$, $\sigma_1=4.4\times10^{-6}$. Initial laser wavelenghts: 800 nm.}
\label{fig:muon}
\end{figure} 

Gaining the ability to produce high-quality muon beams would provide unprecedented capabilities for fundamental-physics experiments in pioneering research topics, ranging from investigations on neutrino-oscillation physics in neutrino-factory machines to energy-frontier studies in TeV-level muon-antimuon colliders \cite{NFMC1,NFMC2,NFMC3,NFMC4}. 
From the point of view of technology, muon acceleration is an extremely challenging task, involving a series of delicate processes, namely, producing high-energy muons via interaction of intense proton beams with appropriate targets, cooling the muon population in order to lower the emittance to acceptable levels (proposed methods are ionization cooling \cite{ioniz_cool_1,ioniz_cool_2,ioniz_cool_3,ioniz_cool_4,ioniz_cool_5,ioniz_cool_6} and frictional cooling \cite{Frictional_1,Frictional_2,Frictional_3}), and, finally, accelerating cold muons to high energy. Moreover, due to the short muon lifetime (2.2 $\mu$s), these processes must be performed rapidly to minimize decay losses.

\begin{figure}[!htbp]
\centering \epsfig{file=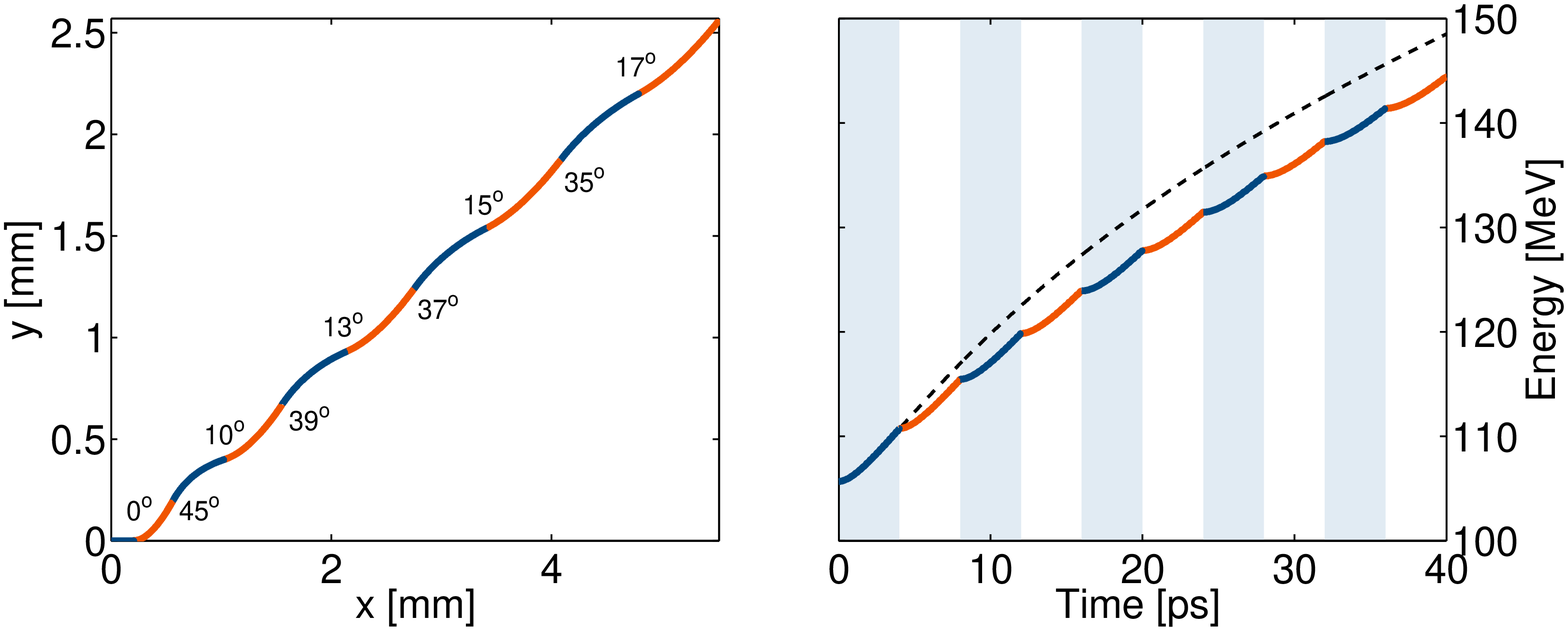, width=4in}
\caption{Multi-stage muon acceleration: (left panel) muon trajectory in the $x$-$y$ plane and (right panel) evolution of the muon energy (solid line) during multi-stage acceleration. The duration of each stage is for 4 ps. In each stage $s$, the laser propagation direction is perpendicular to the muon momentum at the entrance of the stage, ${\mathbf p}_{0s}$ (the values of the angle between ${\mathbf p}_{0s}$ and the ${\mathbf e}_{x}$ are indicated).  Laser 1 is linearly chirped with $\sigma_{1s} = 5.47\times10^{-5}/\sqrt{1+\hat{p}_{0s}^2}$, while the frequency of laser 2 is fixed; the laser intensities are $I_1=I_2=5\times10^{18} \ \mathrm{W}/\mathrm{cm}^2$. In the right panel, the dashed line refers to the evolution of muon energy corresponding to a single-stage process with $\sigma_{1} = 5.47\times10^{-5}$ and same laser intensities.}
\label{fig:stages}
\end{figure}

The all-optical acceleration scheme described here, as well as other conceivable laser-based techniques, might be appealing because of its potential for ultrafast acceleration, which could be particularly important in the first stages of muon acceleration, especially during, or immediately following, the cooling process.
In the example shown in Fig. \ref{fig:muon}, a test muon is accelerated from zero velocity to $0.9 c$ in just 25 ps over 4 mm in a single stage. Unfortunately, such an acceleration pattern would require very large frequency excursions, which are extremely difficult to obtain, at least with current laser technology. A possible strategy to circumvent this obstacle is dividing the acceleration process in several stages such that the acceleration direction in each stage is always perpendicular to the propagation direction of the beam at the entrance of the stage \cite{Gianni}. Theoretically, this would permit the achievement of arbitrary large $\Delta\beta$, by repeatedly using a limited $\Delta\omega$. An example of such a multi-stage acceleration pattern is depicted in Fig. \ref{fig:stages} where moderate-intensity lasers and 10 stages with frequency variations on the order of 50 \% are used to accelerate a muon from zero-velocity to $0.7c$.

Finally, another application of the present method might be the prompt extraction of low-temperature muons from a gas cell used for frictional cooling \cite{Frictional_1,Frictional_2,Frictional_3}, in which the muon distribution relaxes to thermal equilibrium with energy below the ionization threshold of the gas atoms. A possible strategy could be trapping the muons in the beat-wave structure generated by non-ionizing (e.g., with high power and long wavelength) EM waves. The assessment of the possibility of actually meeting the necessary combination of EM-wave parameters (i.e. wavelength, intensities, and chirp law) will be object of future investigation.      

\section{Conclusions}
\label{sec:conclusions}
All-optical trapping and acceleration of heavy particles in vacuum can be achieved by employing counterpropagating, intense laser beams with variable frequencies, which also offers unique control capabilities on important beam features such as the final energy and energy spread \cite{Peano_NJP,Peano_IEEE}. Allowing for ultrafast acceleration, the technique could provide a route towards prompt muon acceleration in the first acceleration stages in a muon collider or a neutrino-factory machine.  

\ack
Work partially supported by FCT (Portugal) through grants PDCT/POCI/66823/2006 and SFRH/BD/22059/2005, and by the European Community - New and Emerging Science and Technology Activity under the FP6 "Structuring the European Research Area" programme (project EuroLEAP, contract number 028514). FP and RB would like to thank the STFC Centre for Fundamental Physics for support.

\end{document}